\documentclass[cameraready]{Interspeech}

\usepackage{comment}
\usepackage{cite}
\usepackage{tipa}
\usepackage{makecell}
\usepackage{amsmath,amssymb,amsfonts}
\usepackage{algorithmic}
\usepackage{graphicx}
\usepackage{textcomp}
\usepackage{xcolor}
\usepackage{tipa}
\title{Towards Language-Agnostic Speech Inversion }

\author[affiliation={1}]{Saba}{Tabatabaee}
\author[affiliation={2}]{Mark}{Tiede}
\author[affiliation={3}]{Suzanne}{Boyce}
\author[affiliation={4}]{Liran}{Oren}
\author[affiliation={1}]{Carol}{Espy-Wilson}

\address{
    $^1$ Department of Electrical and Computer Engineering, University of Maryland College Park\\
    $^2$ Department of Psychiatry, Yale University \\
    $^3$ Department of Communication Sciences and Disorders, University of Cincinnati\\
    $^4$ Department of Otolaryngology-Head and Neck Surgery, University of Cincinnati
}

\email{sabatb@umd.edu, mark.tiede@yale.edu, boycese@ucmail.uc.edu, orenl@ucmail.uc.edu, espy@umd.edu}


\usepackage{comment}

\begin{document}

\maketitle

\begin{abstract}
Characteristic timing patterns are reflected in the acoustic speech signal, encompassing both vocal tract configuration and acoustic excitation. Previous studies have demonstrated that speech inversion (SI) systems can recover these timing patterns from speech, including oral tract variables (tongue and lip constrictions) and source information such as periodic and aperiodic energies and fundamental frequency. In this study, we develop an SI system that simultaneously estimates oral tract variables and three source information parameters trained on co-recorded American English speech audio and articulatory kinematics and investigate cross-linguistic generalizability by evaluating performance on previously unseen languages. Pearson product–moment correlation scores of 0.83 and 0.74 were achieved on untrained French and Russian respectively, across oral tract variables and source information when comparing estimated data with ground-truth measurements.

\end{abstract}

\section{Introduction}
Speech articulation is a complex process that requires precise, temporally coordinated movements of multiple articulators, including the lips, tongue, jaw, velum, and glottis \cite{stevens2000acoustic}. Direct observation of articulation provides the most reliable evidence
of language-specific temporal patterns. However, collecting articulatory data requires specialized equipment, which is costly,
difficult to operate, often unsuitable for many populations of
interest (such as children), and challenging to deploy in field
settings. Characteristic timing and spatial patterns of articulation are also encoded in the acoustic speech signal, which reflects both the configuration of the vocal tract and its dynamic excitation. Due to acoustic interaction effects, as well as the fact that different vocal tract configurations can produce the same output acoustic patterns, this encoding is not exact \cite{atal1978inversion}. However, acoustic recordings are comparatively easy to obtain and analyze, and they have been widely used to investigate accessible language-specific articulatory patterns reflected in the speech acoustics (e.g., \cite{lisker1964cross}). 

To recover characteristic temporal patterns from speech, a speech inversion (SI) system was developed that maps the acoustic signal not to discrete articulatory positions, but to coordinated articulatory synergies known as vocal tract variables (TVs) \cite{browman1986towards, saltzman1989dynamical}. TV trajectories capture the way time-varying synergistic movements of the speech articulators (lips, tongue tip, tongue body, velum, and glottis) are coordinated to achieve acoustic goals. For example, the lip area TV represents the speech goal of lip approximation, achieved depending on context by differing contributions from the upper and lower lip and jaw.  Oral TVs are estimated from ground-truth observations of supralaryngeal articulator positions obtained from flesh-point tracking systems (e.g., electromagnetic articulography (EMA) and X-ray microbeam).  

Acoustic goals in the oral portion of the vocal tract are captured by oral TVs,
which describe the locations and degrees of constriction of the lips, tongue body, and tongue tip. Specifically, these include: lip aperture (LA), lip protrusion (LP), tongue body constriction location (TBCL), tongue body constriction degree (TBCD), tongue tip constriction location (TTCL), and tongue tip constriction degree (TTCD). In addition to oral sounds, nasality is a key contrastive feature in speech production \cite{fry2004phonics}. The velopharyngeal (VP) TV represents coordinated synergistic movements of the velum and pharyngeal wall at the VP port, whose degree of opening controls airflow and acoustic coupling between the oral and nasal cavities. In this study, the performance of the SI system in estimating VP movement is tested against the ground-truth measure of nasalance. The nasalance measure is defined as the ratio of nasal acoustic energy to the combined nasal and oral acoustic energies. This measure provides an indirect measure of VP activity and serves as an accurate proxy for closing/opening movement of the VP port \cite{siriwardena2024speaker}.   

Previous studies \cite{tabatabaee25b_interspeech, siriwardena2023secret} have shown that incorporating source features (SFs), including periodicity (Per), aperiodicity (Aper), and fundamental frequency (F0) into the training of an SI system as a proxy for glottal control can improve its performance. The study by \cite{tabatabaee25b_interspeech} shows that using SFs alongside oral TVs and VP TV as outputs helps the system more accurately estimate oral and VP TVs. These results are consistent with the well-established principle that articulators operate synergistically. Building on these findings, this study integrates the three SFs, along with oral TVs and VP TV, as outputs in the SI system.


Recent advances in self-supervised learning (SSL)-based speech models, such as HuBERT \cite {hsu2021hubert}, Wav2Vec2 \cite{baevski2020wav2vec} and WavLM \cite{chen2022wavlm}, have shifted the development of SI systems toward more robust feature extraction techniques, surpassing traditional acoustic features such as Mel-frequency cepstral coefficients. Studies \cite{hao2024exploring, cho2023evidence, siriwardena2024speaker} show that speech model–based representations outperform conventional acoustic features in SI tasks and generalize better across multiple corpora \cite{siriwardena2024speaker}. Among these models, WavLM-Large has been shown \cite{cho2023evidence,tabatabaee2025acoustic} to outperform HuBERT-Large, as well as other models such as Wav2Vec2 in SI performance. Accordingly, this
work employs WavLM-Large to extract speech representations. 
 
Previous SI systems for estimating oral TVs and VP TV have largely been developed and evaluated using English-language datasets \cite{sun2022unsupervised, tabatabaee25b_interspeech, siriwardena2023secret, wu2023speaker, tabatabaee2026towards}. However, because vocal tract acoustics are shared across humans, much of what SI systems do should be language-independent. To date, 
cross-lingual validation has received limited attention for estimating TVs  \cite{wieling2017analysis}. Therefore, in this study we evaluate the ability of an SI system trained on English, a high-resource language, to recover TVs from speech in other languages. We examine the system’s performance on individual EMA sensor trajectories using example data from English, French, and Russian. Moreover, given the importance of nasality in speech production, we extend the evaluation by assessing how well the SI system estimates the VP TV across languages. To the best of our knowledge, this work presents the first multi-lingual evaluation of an SI system for estimating oral TVs, VP TV, and SFs.

\textbf{Our key contributions} in the current work are as follows:
\begin{itemize}
    \item Collecting a multi-lingual dataset comprising co-collected EMA, nasalance, and speech audio data.
    \item Performing a cross-lingual evaluation of an SI system for the estimation of oral TVs and SFs.
    \item Extending the evaluation of the SI system for VP TV estimation across languages.
    \item Comparing the performance of the proposed SI system for oral TVs and SFs estimation with a recent work in \cite{tabatabaee25b_interspeech}.
\end{itemize}



\section{Methodology}
\subsection{Dataset description and pre-processing}

\subsubsection{XRMB dataset}
The University of Wisconsin X-Ray Microbeam (XRMB) dataset \cite{westbury1994speech} provides speech recordings accompanied by articulatory trajectory data collected using a rasterized X-ray microbeam system tracking pellets placed midsagittally on the oral articulators. The dataset includes recordings from 46 native English speakers (21 males and 25 females), totaling 7,285 utterances and approximately 5.74 hours of speech. The dataset is partitioned into speaker-independent sets for training, development, and testing, as detailed in Table 1. The original pellet coordinates were mapped to TVs using a geometric transformation as described in \cite{attia2024improving}. Since TVs are defined as relative measures (vocal tract constriction location and degree), this mapping minimizes variability arising from anatomical differences among speakers. Following transformation, the processed XRMB dataset includes six oral TVs: LA, LP, TBCL, TBCD, TTCL, and TTCD. All generated TV trajectories were sampled at 100 Hz and normalized to the range of -1 to 1. 
\begin{table}[ht!]
\caption{Description of the XRMB dataset with speaker-independent splits.}
\centering 
\resizebox{\columnwidth}{!}{
    \begin{tabular}{|l|c|c|c|c|}
\hline
Split & Language & \# Subjects & Duration (Min) & \# Utterance \\
\hline
 Train &English & 36& 268& 5801  \\
 \hline
  Development & English & 5& 37&  730 \\
  \hline
   Test &English & 5&38&  754 \\
 \hline
    \end{tabular}
    \vspace{-3mm}
       }
\end{table}
\vspace{-3mm}
\subsubsection{YU dataset}
The XRMB dataset includes only lip and tongue ground-truth data. To enable estimation of a VP TV, we simultaneously collected EMA measurements for oral TVs and nasalance data following the method in \cite{tabatabaee25b_interspeech}. We refer to this dataset as the YU dataset. The YU dataset includes EMA measurements, nasalance, Electroglottography (EGG) signals, and audio recorded via a microphone placed about one foot from participants. Including YU dataset enabled us to extend the SI system to estimate the VP TV, as well as oral TVs and SFs. Speech data were collected from 27 subjects: 20 English, 4 French, and 3 Russian speakers, resulting in 7,451 utterances totaling 7.17 hours. The dataset was divided into training, development, and test sets in a speaker-independent manner (see Table 2 for details). Sensors were placed midsagittally on the tongue tip, tongue blade, tongue dorsum, upper and lower lips, and lower incisors, together with reference sensors used to correct for head movement. Following the geometric transformation to tract variables, the YU dataset included the same TVs obtained for the XRMB data. All TVs were resampled at 100 Hz and normalized to lie within the range -1 to 1. 
\begin{table}[ht!]
\caption{The YU dataset description with speaker-independent splits.} 
\centering 
\label{speaker_verification_models}
\resizebox{\columnwidth}{!}{
    \begin{tabular}{|l|c|c|c|c|}
        \hline
        Split &Language& \# Subjects  & Duration (Min) & \# Utterance \\
        \hline
        Train & English& 12  & 193 & 3230\\
        \hline
        Development & English& 4  & 68 & 1235\\
        \hline
        Test & \makecell{ English\\French \\  Russian}  & \makecell{4 \\ 4 \\  3}  &  \makecell{ 75 \\59 \\  35} & \makecell{ 1170 \\1293\\ 523 }\\    
        \hline
    \end{tabular}
    \vspace{-3mm}
       }
\end{table}


\vspace{-3mm}
\subsection{Developing SI system}
The SI system was developed using a multi-task learning framework. The first task estimates six oral TVs, while the second task estimates the three SFs, enabling simultaneous estimation of all parameters. The SI system was trained on the combination of the XRMB and YU English datasets, as summarized in Table 1 and 2. Ground-truth values for the three SFs in XRMB and YU datasets were extracted from speech signals using the APP detector \cite{deshmukh2005use}.
\begin{figure}[htbp]
    \hfill
  
    \includegraphics[width=0.49\textwidth, height=0.25\textheight]{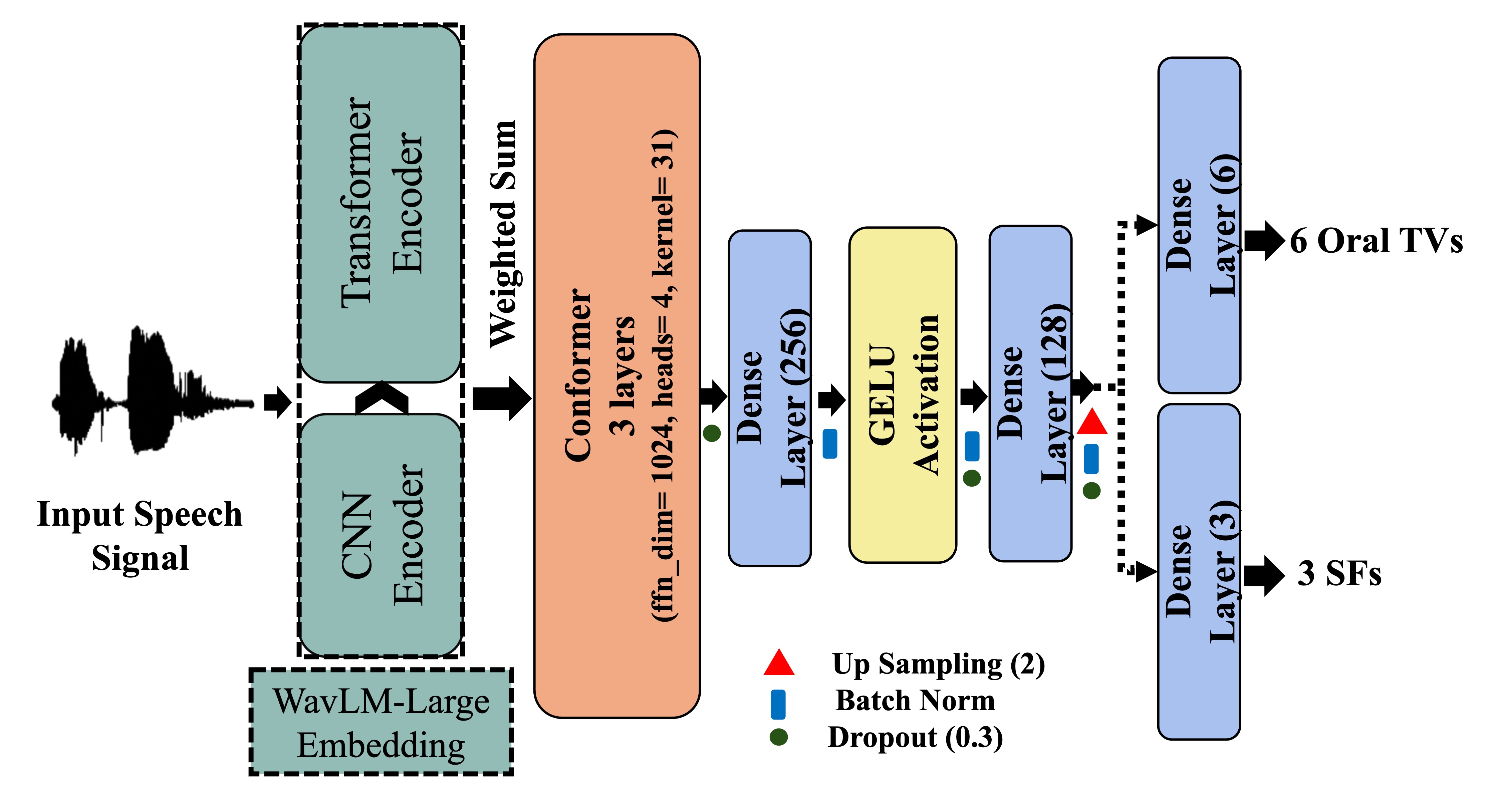}  
    
    \caption{Proposed model architecture for the SI system.}    
    \label{fig:si}
    
\end{figure}

\begin{table*}[htbp]
\caption{PPMC scores for the estimated parameters of the SI model. Avg all: average PPMC score computed across all nine parameters.}
\setlength{\tabcolsep}{6.5pt}
\renewcommand{\arraystretch}{1}
\centering
\begin{tabular}{l|l|ccccccccc|c}
\hline
 Model &Language  &LA & LP & TBCL & TBCD & TTCL & TTCD & Per & Aper & F0 &Avg all \\
\hline
\multicolumn{10}{c}{\hspace{3cm} \textbf{XRMB test set}} \\ \hline

SI model in \cite{tabatabaee25b_interspeech}& English  & 0.91&0.76&0.80&0.86&0.84&0.95  &0.94 & 0.88 & 0.75 & 0.85\\
Proposed SI model&English  & 0.93&0.76&0.78&0.87&0.82&0.95  &0.95 & 0.90 & 0.79 & 0.86\\
\multicolumn{10}{c}{\hspace{3cm}  \textbf{YU test sets} } \\ \hline

Proposed SI model&English   &0.87 & 0.86& 0.81 & 0.84 & 0.82 & 0.87  & 0.95 &0.83 & 0.84 &0.85\\ 


Proposed SI model&French &0.88 &0.87 & 0.80 & 0.83 &0.78 & 0.84  & 0.95 & 0.80 & 0.76 &0.83\\

Proposed SI model&Russian &0.76 & 0.75 & 0.70 & 0.71 & 0.69 & 0.71 & 0.91 & 0.68 & 0.71 & 0.74 \\ 
    \end{tabular}
    \vspace{-3mm}
\end{table*}
\vspace{-3mm}
As illustrated in Figure 1, we employed the pre-trained WavLM-Large model to extract speech representations from the input signals. Representations from all 25 hidden layers of WavLM-Large were concatenated, and a weighted sum of these layer-wise embeddings was computed to produce a single, unified representation. This combined representation was subsequently processed through three Conformer layers, which capture both local and long-range temporal dependencies in the speech signal. The Conformer output was passed through a fully connected layer with 256 hidden units, followed by a Gaussian Error Linear Unit (GELU) activation. This was followed by a second fully connected layer consisting of 128 hidden units. To account for the mismatch in temporal resolution between the WavLM embeddings (sampled at 50 Hz) and the target outputs (sampled at 100 Hz), we applied upsampling by a factor of two, ensuring alignment with the target sampling rate. Finally, multi-task learning was implemented through two separate output layers: one dense layer with six units for estimating the six oral TVs, and a second dense layer with three units for predicting the three SFs. 

Training used the AdamW optimizer with an initial learning rate of 5e-4 weight decay of 1e-3, and a batch size of 8. A plateau-based scheduler reduced the learning rate when validation performance failed to improve for five consecutive epochs. Early stopping with a patience of eight epochs was applied to mitigate overfitting. Task-specific losses in the SI system were computed using Equation~\ref{eq:loss_function}, and the total loss was defined as the sum of all task losses. The loss function combines Pearson correlation (PC) and root mean square error (RMSE), with $\alpha = 0.2$ empirically chosen for optimal performance.
\begin{equation}
\mathcal{L} = (1 - \text{PC}) + \alpha \, \text{RMSE}
\label{eq:loss_function}
\end{equation}

\section{Results and discussion}

\subsection{Performance of the SI system}
Table 3 summarizes the performance of the proposed SI system on the XRMB and YU test sets, evaluated using Pearson product–moment correlation (PPMC) scores, which measure the alignment between estimated and ground-truth articulatory parameters. On the XRMB test set, the proposed SI system achieves an average PPMC of 0.86 across all nine articulatory parameters. This result slightly exceeds the average PPMC of 0.85 reported in the previous work \cite{tabatabaee25b_interspeech}, which used the same XRMB geometric transformations and train-test partitioning. For the YU English test set, the system achieves an average PPMC score of 0.85, showing its ability to estimate articulatory parameters for the language on which it was trained. Collectively, these results demonstrate the SI system’s ability to estimate oral TVs and SFs across both the XRMB and YU English datasets.

To evaluate cross-lingual generalization, the SI system was further tested on the French and Russian subsets of the YU dataset. Notably, even though these languages were not seen during training, the system continues to demonstrate strong performance, achieving average PPMC scores of 0.83 for French and 0.74 for Russian across all nine parameters. The SI system estimate for pitch and periodicity across all Russian speakers was lower than that for French speakers.  Our analysis suggests that this was primarily due to one Russian speaker being recorded in a noisier environment.

Overall, these results highlight the SI system’s ability to generalize across languages, performing well on the training language while retaining strong predictive capability on previously unseen languages. Performance on French and Russian indicates that the system captures shared underlying articulatory patterns, underscoring its potential for multi-lingual speech modeling and cross-linguistic articulatory research.

\subsection{Comparison of estimated and ground-truth oral TVs}
Figure 2 shows the estimated TBCD, TTCD, and LA TVs produced by the SI system, together with the corresponding ground-truth measurements, for randomly selected English, Russian and French utterances from the YU test set. A detailed comparison between the SI system estimates and the ground-truth values shows that, despite being trained exclusively on English speech, the SI system effectively captures the articulatory patterns of all three TVs across English, Russian and French languages, demonstrating its ability to model fine-grained articulatory behaviors in previously unseen languages and underscoring its potential applicability in multi-lingual speech processing and articulatory modeling.

\begin{figure}[htbp]
    \hfill
    \includegraphics[width=0.52\textwidth, height=0.45\textheight]{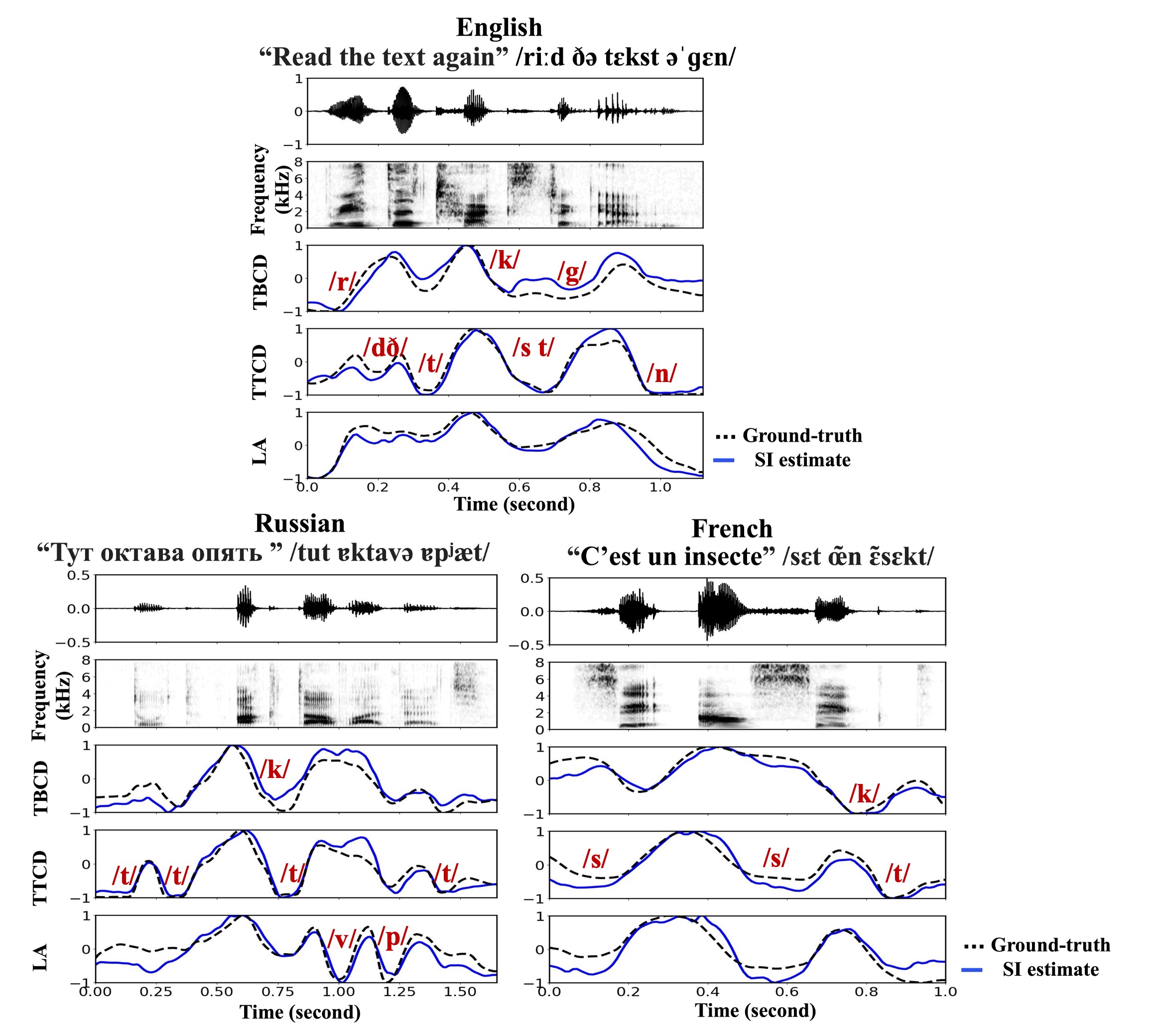}  
    \caption{Waveforms and spectrograms of English, Russian and French utterances, with comparisons between ground-truth TBCD, TTCD, and LA, and the corresponding estimates from the SI system.}    
    \label{fig:si}
    \vspace{-4mm}
\end{figure}
\subsection{Performance of the SI system for VP TV estimation}
To further evaluate the cross-linguistic capability of the SI system, we extended the analysis beyond oral TVs and SFs to include estimation of a VP TV. Specifically, we investigated whether an SI system trained exclusively on English speech data can estimate the VP TV for other languages. This is particularly interesting because English and French have very different articulatory patterns of nasalization; French has a phonemic distinction between oral and nasal vowels while nasalization is not distinctive for vowels in English.   

For the development of the SI system, we utilized the XRMB and YU English datasets. The YU dataset includes ground-truth nasalance measurements, whereas the XRMB dataset does not provide nasalance ground-truth for its speech signals. To address this limitation, we employed the nasal SI system proposed in \cite{tabatabaee25b_interspeech} to estimate nasalance ground-truth values for the XRMB dataset from the speech signals, effectively retrofitting XRMB with nasalance measurements. This approach is inspired by \cite{tabatabaee25b_interspeech}, which demonstrated that nasalance estimates generated by a nasal SI system for the XRMB dataset can be used to train an SI system effectively, as validated by comparison with a real ground-truth nasalance dataset.

To incorporate the VP TV as an additional output, the SI system’s output architecture was modified. Specifically, the original final dense layer of three units (used for estimating the three SFs) was replaced with a dense layer of four units, enabling simultaneous estimation of the three SFs and the VP TV (see Figure 1). Following a similar multi-task framework, one final layer estimates six oral TVs, while another dense layer predicts the three SFs along with a VP TV. The SI system was trained using the XRMB and YU English datasets, as summarized in Tables 1 and 2. The trained model was subsequently evaluated for nasalance estimation on the YU English test set, as well as on three French speakers and one Russian speaker for whom ground-truth nasalance measurements were available. 

As shown in Table 4, the PPMC scores between the VP TV estimates and the nasalance ground-truth values are high for English speakers, reaching 0.92 on the YU dataset. Moreover, the system also achieves strong PPMC scores for languages that were not seen during training, including Russian, and French. These results demonstrate the potential of the SI system for reflecting language-specific patterns without the need for language-specific training.

\begin{table}[htbp]
\vspace{-1mm} 
\caption{PPMC scores comparing the VP TV estimated by the SI model with nasalance ground-truth on the YU dataset.}
\footnotesize
\setlength{\tabcolsep}{7pt}
\renewcommand{\arraystretch}{1}
\centering
\begin{tabular}{c|c|c}
\hline
 Language &  \# Subjects &  VP     \\
\hline
 
English &4 &0.92\\
French &3 &0.89\\
Russian &1 &0.82\\ \hline

    \end{tabular}
    \vspace{-2mm}
      \end{table}
\vspace{-3mm}
\subsubsection{Comparison of estimated and ground-truth nasalance}
Figure 3 shows a randomly selected example from the XRMB test set, illustrating the nasalance ground-truth and the SI system estimates for the utterances “It's seam ore Sid” and “It's see more Sid”. 
\begin{figure}[htbp]
    \hfill
    \vspace{-3mm}
    \includegraphics[width=0.5\textwidth, height=0.2\textheight]{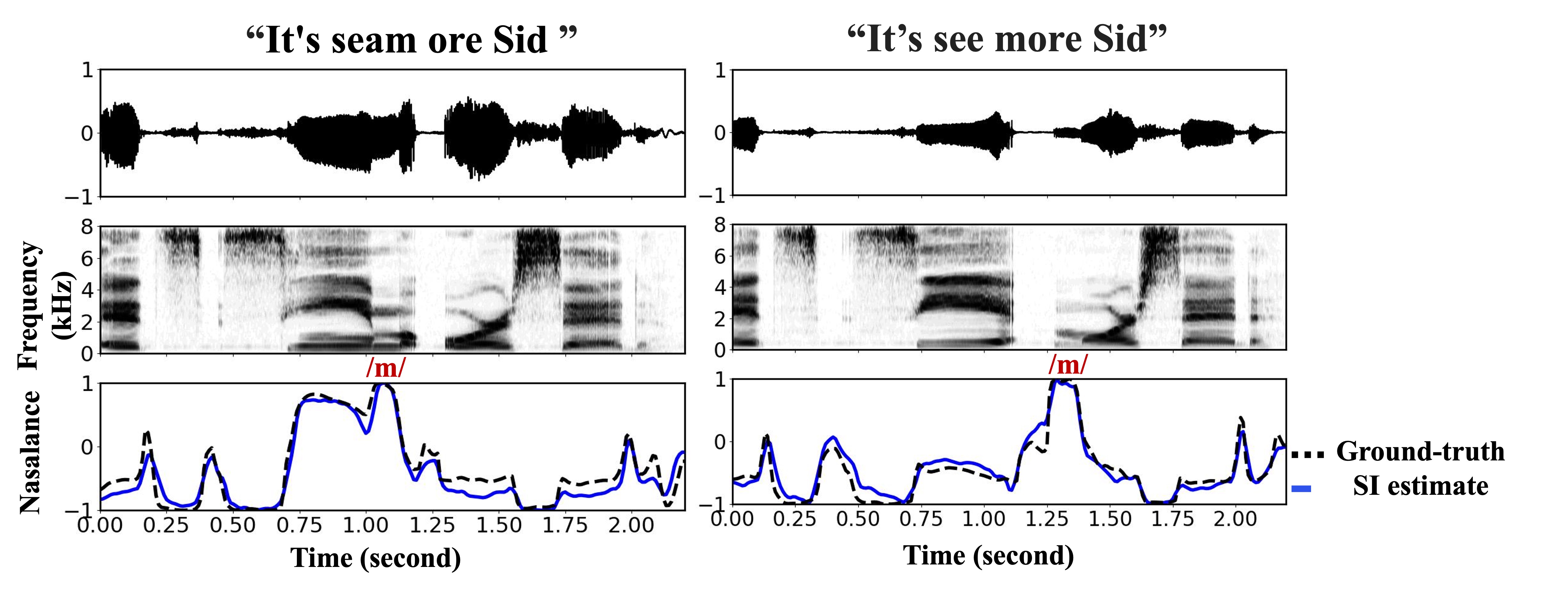}  
     \vspace{-8mm}
    \caption{Waveforms and spectrograms of the utterances “It's seam ore Sid” and “It's see more Sid”, with corresponding comparisons of ground-truth nasalance and the VP TV estimated by the SI system.}    
    \label{fig:si}
    \vspace{-2mm} 
\end{figure}
For the instances of the phoneme /m/ in “seam” and “more” the estimated nasalance exhibits a peak in each case, which is consistent with the nasalance ground-truth. In the utterance “It's seam ore Sid”, the nasalance peak begins well before the production of /m/, indicating that the VP port starts to open prior to the articulation of the nasal /m/ sound, thereby producing anticipatory nasalization. In contrast, in “It's see more Sid”, the presence of a word boundary blocks anticipatory nasalization. Figure 4 presents a comparison between the VP TV estimates generated by the SI system and the corresponding ground-truth nasalance values for randomly selected samples from the YU test set for French and Russian. In the French example, nasalization occurs in an oral-nasal sequence involving /n/, as in “bonne”. In the Russian example, nasalization occurs on the nasal consonants /m/ and /n/ within the utterance. Across all examples, the SI system VP TV estimate closely follows the ground-truth nasalance.

\begin{figure}[htbp]
    \hfill
    \vspace{-3mm}
    \includegraphics[width=0.51\textwidth, height=0.2\textheight]{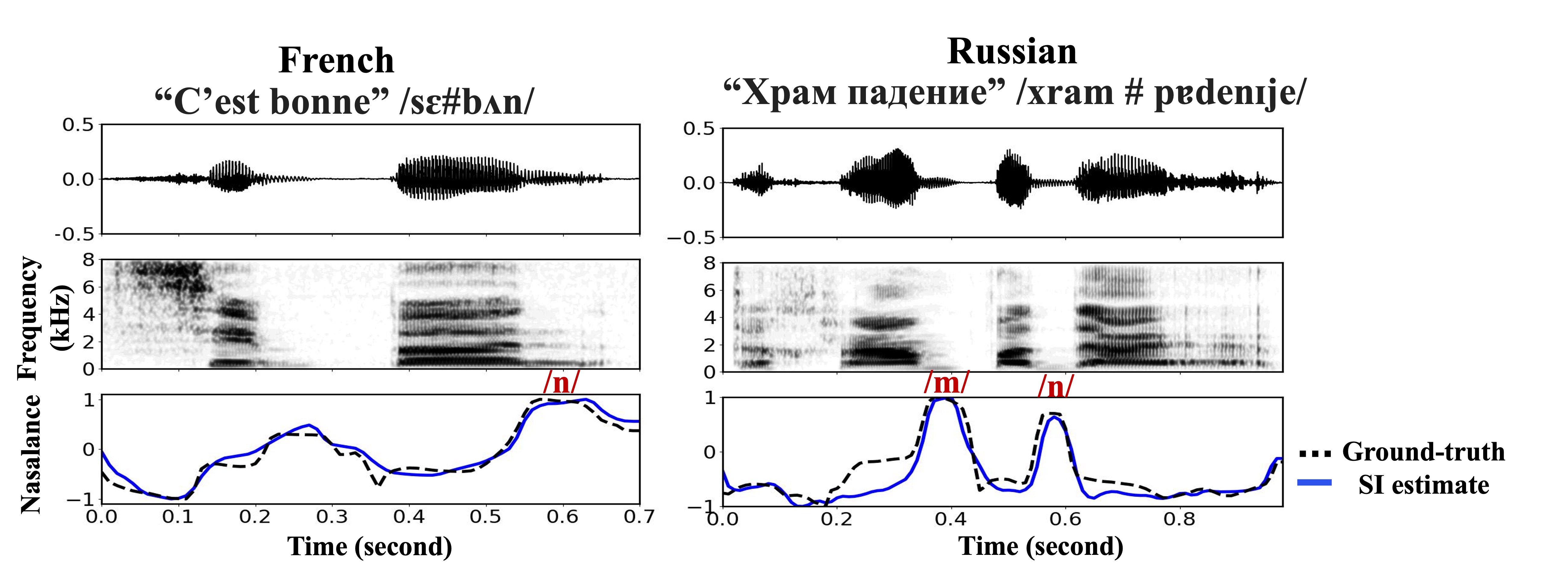}  
     \vspace{-8mm}
    \caption{Waveforms and spectrograms of French and Russian utterances, alongside comparisons between ground-truth nasalance and the VP TV estimated by the SI system.}    
    \label{fig:si}
    \vspace{-2mm} 
\end{figure}
     \vspace{-3mm} 
\section{Conclusions and future work}
In this paper, we investigated the performance of an SI system trained on English data when applied to non-English languages, including French and Russian. We developed an SI system capable of simultaneously estimating oral tract variables and three source features, and evaluated its performance on both English and non-English speech. The results demonstrate that an SI system trained on English can be effectively applied to other languages such as French and Russian for oral tract variable trajectory and source information estimation from the input audio signal. Furthermore, we show that the SI system is capable of estimating velopharyngeal opening/closing patterns across languages, highlighting its cross-lingual generalizability. By jointly estimating multiple speech parameters within a unified framework and showing strong generalization to unseen languages, the proposed SI system demonstrates potential for multi-lingual speech processing and clinical assessment applications. Future work will focus on collecting data from more speakers and evaluating the proposed SI system on additional languages, particularly under-resourced, languages.
\section{Generative AI Use Disclosure}
Generative AI tools were used solely for spelling and grammar correction.
\bibliographystyle{IEEEtran}
\bibliography{mybib}

\end{document}